\documentclass{iau}
\usepackage{graphicx}

\title[Stars and exoplanets in Stokes IQUV] 
{Stars and exoplanets in Stokes IQUV: a decadal opportunity for HIRES at the ELT}

\author[K. G. Strassmeier]   
{Klaus G. Strassmeier
}

\affiliation{Leibniz-Institute for Astrophysics Potsdam (AIP), An der Sternwarte 16, D-14482 Potsdam, Germany \\email: {\tt kstrassmeier@aip.de}}

\pubyear{2018}
\volume{347}  
\jname{Early Science with ELTs}
\editors{Norbert Przybilla, Karen Pollard \& Annalisa Calamida, eds.}
\begin{document}

\maketitle

\begin{abstract}
We proposed that the European ELT will be equipped with an ultra-high-precision polarimetric light feed as part of its high resolution optical and near-IR spectrograph HIRES. Such a feed is unique among the new ELTs and only possible in a rotationally symmetric focus of the telescope. The ELT's f/4.4 intermediate focus near M4 could provide such a capability with a polarimetric sensitivity of down to $10^{-5}$  for the brightest targets. Among the new science steps forward with HIRES-Pol at the 39\,m ELT would be the full characterization of solar-like stellar magnetospheres by means of Zeeman-Doppler Imaging. In particular for planet-hosting stars it could constrain the habitability of a planet based on its particle-emission geometry from the host star. Besides, any stellar linear-polarization spectrum is new territory for astrophysics and I refresh reasons why this can be also important for exoplanet atmospheres.
\keywords{polarization, magnetic fields, planetary systems, instrumentation, astrobiology}
\end{abstract}

\firstsection 
\section{Introduction and aim of this paper}

The scope of this talk/paper is to demonstrate that the scientific return of the presently planned two spectrographs of HIRES (VIS and NIR module) could be substantially increased when combined with the power of high-precision spectropolarimetry. It even has the potential to contribute to the search for biosignatures in exoplanet atmospheres, comets, and the interstellar medium.

The majority of the HIRES polarimetry science cases are based on high spectral resolution, high accuracy and very broad wavelength coverage. Even with the ELT, sometimes only the co-addition of many spectral lines with similar polarization properties will reach satisfactory levels of significance. For line polarization, the very broad wavelength coverage is mandatory to reach the highest possible accuracy of any single observation. Accuracy is key because it is the quantitative analysis of physical conditions that lead to the polarization of radiation, where the superiority of spectropolarimetry is largest. No wavelength is clearly preferred but the quadratic wavelength dependency of some magnetic processes favor the near infrared. All of the presented science cases involve unresolved sources.

The distribution of preferred spectral resolutions is fairly flat from low ($R=\lambda/\Delta\lambda\approx 10,000$) to very high ($R\approx100,000+$). The accuracy demanded is always high and should be 0.1-0.5\% per spectral resolution element and observation in the degree of the polarization. This shall lead to an achievable sensitivity of $10^{-5} S/I$, where $S$ is any of the Stokes parameters QUV and $I$ the intensity. Even for the Sun, this is only achievable after co-adding individual observations. However, high spectral resolution spectropolarimeters with large wavelength ranges like PEPSI at the LBT (Strassmeier et al. \cite{pepsi}) and HIRES at the ELT will allow the use of thousands of spectral lines to boost the signal-to-noise ratio (S/N) of an average spectral line by a factor 100 or more (five orders of magnitude in brightness range). Of course, for the cost of ``averaging''.

The next sections are excerpts from the science cases document for spectropolarimetry with eltHIRES and are by no means complete. I focus here on cool stars with a magnetosphere and exoplanets. Send a mail to the author for a pdf copy of the full document.

\section{Application to stars}

\subsection{Recalling solar polarimetric phenomenology}

For most of the solar disk the degree of linear polarization of the continuum is under 0.1\,\%, but it rises to 1\,\% at the limb. The polarization also depends strongly on wavelength; for near ultraviolet 3000\,\AA\ the light near the limb is 100 times more polarized than red light at 7000\,\AA . Linear polarization in spectral lines is usually created by anisotropic scattering of photons on atoms and ions which can themselves be polarized by this interaction. Circular polarization on the other hand is mainly due to transmission and absorption effects in magnetic surface regions. Circular line polarization scales with the Land\'e factor of the spectral transition, the wavelength, and the magnetic field density. At least on the Sun, there is almost no circular continuum polarization.

The linearly polarized solar spectrum (dubbed the second solar spectrum) differs significantly from the solar spectrum determined by the intensity of light (Gandorfer et al. \cite{gan}, Stenflo \& Keller \cite{ste:kel}). Large effects come around the Ca\,{\sc ii} H \& K line but very weak lines in the intensity spectrum are able to produce large polarization signatures. Molecular lines with stronger polarization than the background due to MgH and C2 are common. Rare Earth elements stand out far more than expected from the intensity spectrum. Other odd lines include Li\,{\sc i} at 6708\,\AA\ which has 0.005\% more polarization at its peak, but is almost unobservable in the intensity spectrum (Stenflo \cite{ste11}). The Ba\,{\sc ii} 4554\,\AA\ line (c.f., Belluzzi et al. \cite{bellu}) appears as a triplet in the second solar spectrum but is a single line in the intensity spectrum. This is due to differing isotopes and hyperfine structure. The two D1 lines of sodium and barium at 5896\,\AA\ and 4934\,\AA, respectively, were predicted by quantum theory not to be polarized, but nevertheless are present in the solar spectrum.

There exists no linearly polarized spectrum for any other star than the Sun with reasonable S/N ratio. The simple reason is the lack of photons. A well exposed spectrum of any other solar-like star would already be a unique milestone-like achievement for HIRES and the ELT.

\subsection{Towards full stellar characterization}

Once a star has been discovered to host another Earth within the habitable zone, the race is on to characterize all facets of the host star. Quantitative questions of stellar differential rotation, wind geometry, and stellar magnetic fields will appear. Most obvious, a full characterization must provide information on the stellar magnetic field and its 3D geometry. This can only be provided from a full Stokes vector IQUV spectrum at high spectral resolution and as a function of rotational phase of the star. Only then one can invert the polarized line profile changes due to surface inhomogeneities into a surface magnetic field together with its morphology. This powerful technique is called Zeeman-Doppler imaging (ZDI; e.g., Donati \& Brown \cite{don:bro}; Piskunov \& Kochukhov \cite{pis:koc}; Carroll et al. \cite{car:str}) and is able to provide the 2D surface information. Its extrapolation out into the planetary environment may hold evidence for the existence of a planetary magnetic field (open stellar field lines might indicate the position of the planet). Stellar and planetary magnetism are a prerequisite for the evolution of complex organic molecules because of its protective character against high-energy cosmic rays.

We have seen large consortia to follow up CoRoT, Kepler and K2 discoveries and this will also be the case for TESS and later PLATO (launch 2026), complemented by ground-based surveys. Among the currently most interesting targets are Proxima Centauri (Anglada-Escud\'e et al. \cite{angl}; followed by 50+ papers on this target in 2016 and 2017), Trappist-1 (Gillon et al. \cite{gill}; plus another 22 papers thereafter), or Boyajian's star (Boyajian et al. \cite{boy}; plus another 22 papers in 2016 and early 2017). None of these targets is in reach of current (high-resolution) spectropolarimetric facilities, only a polarimeter for HIRES and the ELT will enable to reach such targets (JWST does not have any polarimetric capability).

\subsection{At the hydrogen-burning limit: magnetic fields in fully convective stars}

There are competing theories on the type of dynamo mechanism operating in fully convective stars. To down-select them, it is necessary to measure differential rotation and the magnetic field configuration of stars from late-M types to the limit of stellar hydrogen burning (see Donati et al. \cite{don:for}). Only the brightest M-type stars can currently be studied in this manner. Recent results suggest that very low mass stars down to ~0.09\,M$_\odot$ possess strong kG fields (Reiners et al. \cite{rein}). However these are just based on Stokes~I observations; analysis with Stokes Q, U, and V is needed for the complete understanding. Objects below the hydrogen-burning limit may possess magnetic fields, but these are very hard if not impossible to detect with 8--10m class telescopes (Kuzmychov et al. \cite{kuz}). A straightforward goal for the ELT would be to detect and then map magnetic fields on the surfaces of L dwarfs near the hydrogen burning limit.

\section{Application to extra-solar planets}

\subsection{Spectropolarimetry as a new tool}

The light-collecting power and angular resolution of very large telescopes become indispensable when measurements of the light reflected or transmitted by an exoplanet are attempted. The task of characterizing exoplanets is a major field of endeavor, with the identification of potentially habitable planets as a key driver for new ground and space-based instrumentation (e.g., Snellen et al. \cite{snell}).

Exciting progress is being made already, giving a glimpse of what the ELT may achieve. For example, in the brightness variations of a giant eclipsing planet, the CoRoT satellite has detected the differences between the illuminated and the dark hemisphere (Snellen et al. \cite{snell09}). The presence of an atmosphere and a value of the albedo were derived (see also Martins et al. \cite{mar18}). This is yet to be compared with albedos from polarimetry. Many more papers followed from Kepler and K2 data. Martins et al. (\cite{mar15}) directly detected the hot Jupiter of 51~Peg in high-resolution optical spectra with HARPS on the 3.6m. By applying the cross-correlation technique where the planet's location in wavelength space is well known from the orbital elements the star-planet intensity ratio was $10^{-4}$, thus comparably large.  Evidently, spectra would reveal incomparably much more physical information than light curves, only to be further surpassed by spectropolarimetry (e.g., Garcia Munoz \cite{garcia}). Spectral resolution will always be important for accurately removing spectral contaminations, mostly due to the Earth atmosphere but also due to intrinsic magnetic stellar activity. Spectral resolution of, say, $R\approx100,000$, allows detecting the Doppler-shifted absorption lines from the planet's atmosphere, enabling the firm detection of molecules if present, as well as determining the orbital velocity of the planet, and with this the mass of the star and the planet. Also, constraints on the temperature profile and gas abundances can be placed.

\subsection{Reflected light from a planet}

Polarized spectral line observations near orbital quadrature may show up with up to $\approx$20\,\%\  polarization due to reflection (e.g., Stam \cite{stam08}). Additionally, coherent scattering induced by a planetary magnetosphere or due to molecular anisotropies, e.g., in water lines due to its molecular bipolarity, will modulate this polarization. Other atmospheric constituents are CO$_2$, CH$_4$, CO, H$_2$ and Na (see, e.g., Sing et al. \cite{sing} for hot-Jupiter exoplanets). Several spectral lines for these constituents appear in the NIR and the red part of the optical spectrum in particular for water, methane and potassium. Strong stellar lines appear, e.g. for HD\,189733, from Ca\,{\sc ii} H\&K or the Ca\,{\sc ii} NIR triplet at 3950\,\AA\ and around 8600\,\AA\, respectively, or the Na\,{\sc i} doublet at 5895\,\AA. Importantly, stellar lines reflected by the planet become Doppler shifted according to the planet's motion and may be detectable with high spectral resolution and a cross-correlation technique. Note that the ratio Q/I peaks further away from the stellar disk (near 90$^\circ$) than just the pure linear polarization (near 60$^\circ$). This is because Stokes~I peaks naturally when the planet's disk is fully illuminated, i.e., around upper conjunction, which means always close to the stellar disk. The polarization signal Q/I would be thus strongest when the radial-velocity separation between planet and star is largest, which is an observational advantage.

Detecting the spectrum of a Jupiter-sized planet orbiting a Sun as in 51~Peg in linearly polarized light appears comparably easy for an ELT. The predicted flux ratio star-to-planet in Q/I will be driven by the quantity $Q(R_p/a)^2$, where $Q$ is the amount of linearly polarized light reflected off the planet, $R_p$ the planet's radius in units of its distance to the host star, $a$. Values of between $10^{-5\dots -6}$ (or 1-10\,ppm) are predicted (ignoring all other uncertainties and difficulties like ISM contamination or stellar-activity cross-talk). For a conservative back-of-the-envelope estimation it thus requires a S/N per resolution element of 200,000:1 for a 3-$\sigma$ detection. While large indeed, the wavelength coverage of eltHIRES from 400--1800\,nm at $R=100,000$ covers at least 10,000 spectral lines for a G2V host star (more for a K2V star, less for a F2V star). Assuming all these lines show the same polarization signal, it then requires spectra with a S/N of 2,000 in the continuum, i.e., for 51~Peg four $\approx$1-min exposures with the ELT or in total 10\,min telescope time. For 55~Cnc's Super-Earth, we expect a signal of 30~ppm in I and 1-3~ppm in Q. Thus, a 10\,hr exposure should yield a Q($\lambda$)-detection of a Super-Earth. If done over a few phase angles one needs an observing project of a few nights.

Reconstructing not only Q($\lambda$) but also Q($\alpha$,$\lambda$), and thus seeing the large-scale and the small-scale scattering properties, will not only constrain the optical thickness of a cloud deck (and with a model even $T(\tau))$ but also the number of scatterers in the planet's atmosphere. It even constrains the geometry of the (planetary) cloud coverage due to its polar versus equatorial scattering properties (Stolker et al. \cite{stol}) or, if the planet is in transit, the background stellar atmosphere dust particles and its coverage (Sengupta \cite{seng}). Spectral resolution is mandatory to remove line contamination (stellar as well as terrestrial) but also to measure precise Doppler shifts and to detect (complex) molecular line patterns from the planet's atmosphere if existent.

\subsection{Search for biosignatures}

In the search for life, time resolved \emph{linear} spectropolarimetry can be used to seek evidence of special features such as a strongly polarized specular reflection, dubbed the glint (e.g., Robinson et al. \cite{rob:enn}), that would arise from the liquid surface of an extrasolar ocean. The ocean glint appears large at crescent phases, allowing it to strongly influence the crescent Earth's reflectivity, if not hampered by clouds. The significance of the ocean glint to boost the polarization signal of Earth has been demonstrated for Earthshine spectra (Sterzik et al. \cite{sterz}) and by detailed 3-D radiative transfer models (Emde et al. \cite{emde}). However, it is important to note that both Rayleigh and cloud scattering have strong polarization signatures on their own (Stam \cite{stam08}), and either of which can overwhelm the glint polarization signature, or that from rainbows from liquid droplets (e.g., Zugger et al. \cite{zugg}).

Precision \emph{circular} spectropolarimetry may offer a direct probe of the presence of microbial photosynthesis or vegetation. The unique homochirality of biological material coupled to the optical activity of biological compounds means that biological matter can influence the polarization of reflected light, circular polarization in particular. Homochirality arises as a consequence of self-replication hence is likely to be generic to all forms of biological life and has the potential to produce a macroscopic signature. Sparks et al. (\cite{spar}) and others have shown that a variety of photosynthetic microbial organisms, as well as macroscopic vegetation, yield a $10^{-4\dots-5}$ signature in their circular polarization spectra. The strongest change of polarization is found in the photosynthetically active spectral regions around the Vegetation Red Edge (VRE) at 700\,nm. Hence, circular polarization spectroscopy, though technically challenging even with an ELT, offers one of the purest and most direct generic biosignatures at our disposal.

\section{A technical solution: status at phase-A level}

\subsection{ELT focus}

Woche et al. (\cite{woche}) and Di~Varano et al. (\cite{div}) presented the optical and mechanical design of an ELT polarimeter in the context of the Phase-A study for HIRES (Marconi et al. \cite{marconi}). Earlier versions were presented in Di~Varano et al. (\cite{divarano}) and Strassmeier  et al. (\cite{str12}). In order to reduce the instrumental polarization and cross-talk to scientifically acceptable values, the optimal position for a polarimeter along the optical path of the telescope is the rotationally symmetric focus. A trade-off-design study showed a $10^5$-times lower instrumental polarization in the ELT intermediate focus (IF) with respect to the ELT Nasmyth focus (NF) basically in agreement with the NF study of Ovelar et al. (\cite{ovelar}). Access to the IF located near the M4 surface vertex is not trivial but can be achieved with a reimaging system and then accessed at a safety distance of at least 300\,mm from M4. The design of a transfer-optics unit for such a location is mostly constrained by having an allowed vignetting area of maximum 5\,arc min. Thus, all design options require a deployable polarimeter in the telescope beam behind M4 and are realized with a mechanical swing arm. The currently favored optical solution is a double Cassegrain system to reimage the IF to the NF. It includes the polarization optics and feeds the other ELT mirrors by redirecting the ordinary and extraordinary beams to the front end (FE) in the standard Nasmyth focus. The FE module comprises then the components for sky derotation, atmospheric dispersion correction (ADC), wavelength splitting in two bands (BVRI, zYJH), field stabilization, conversion to f/20, and dispatching the light into two pairs of fiber bundles to feed the VIS and the NIR HIRES spectrographs simultaneously.

\subsection{Transfer optics}

The currently favored version is an inverted double Cassegrain optics or alternatively an inverted Gregorian optics. It creates an intermediate pseudo-collimated beam within we can place the polarimetric optics. Its beam is then reimaged by another Cassegrain or Gregorian system to the same f-number before M3 as the f-number in the intermediate focus in order to get to the original f-ratio in the Nasmyth focus just like if the polarimeter would have not been in the telescope beam at all. The field of view of the polarimeter in combination with the inserted transfer optics is reduced to a few arc seconds caused by the strong power of the transfer optics. The imaging is diffraction limited in 3 arc sec FOV which is sufficient for point-source seeing-limited single object observations. Obscuration starts for fields larger than 4\,arc sec. This represents the technologically built-in seeing limit up to within the polarimeter should be used.

\subsection{Polarimetric optics}

The polarization optic is located shortly before the focus in the convergent f/13 beam of the Cassegrain-like first transfer optic. The optical elements are a $\lambda$/4 phase plate as retarder and a double Wollaston prism (DWP) as the polarizer. The DWP splits the beam in two orthogonal polarized components and the following CaF$_2$ singlet reduces the divergence of the separated beams to feed the telescope transfer optics properly. The image quality of the transfer optics is reduced by the beam divergence in the DWP and by chromatic dispersion in both separated beams. Because we transfer the polarized light to the Nasmyth focus, we can ignore the small beam separation at the polarimetric focus and correct for chromatic dispersion in the regular HIRES FE in the Nasmyth focus where there is enough space. Nevertheless, we expect seeing limited spots of less than 0.25\,arc sec from a full Zemax solution. Atmospheric dispersion is not compensated before the polarimetric optics but, again, can be corrected at the Nasmyth focus in the FE. The calibration of the polarimeter can be realized with an insertable and turnable Glan-Thompson prism between Cass-M1 and Cass-M2 and a fiber link that brings calibration light to the intermediate focus. Retarder and Wollaston prisms are turnable as well.

\subsection{Nasmyth front end}

After the two polarized beams arrived at the NF the first thing to do is to derotate their North-South axis from the telescope motion. This can be achieved with a Dove prism or actually a mechanical derotation like in any other image derotator. The Dove-prism solution also acts as a subsequent beam splitter. The separation at the NF between the ordinary (o) vs. the extra-ordinary (eo) beam is 3.6\,arc~sec while a prism after the Dove redirects the o- and ao-beams into 180$^\circ$ opposite directions. Two collimators convert their f/20 beams into two parallel beams within the dichroic separation, the field stabilization, and the ADC is implemented. Finally, four independent optical cameras reimage the entrance aperture onto four fiber bundles that connect to the two spectrographs.

\section{Summary}

We emphasized the importance of spectropolarimetry for stellar and exoplanetary physics. Key science case for cool planet-hosting stars is the capability of a full characterization of the host star and its environment including the stellar magnetosphere. The latter may turn out to be of crucial importance for habitability, as it is the case in our solar system. Zillions of other science cases for stars in general exist but were not mentioned in this contribution. I refer again to the HIRES polarimetric science case document. Key science for exoplanets is the detection of the polarimetric albedo from reflected light. It carries information on the planetary atmosphere absorbers and scatterers. A more visionary goal is the detection of circularly polarized light around the VRE from light transmitted through the planetary atmosphere (during transit). It would be a direct bio signature.

The current design of the polarimeter is such that it would deliver two spectra (from the o- and ao-beams) to the VIS and the NIR modules simultaneously for one predetermined spectral resolution. Its spectral coverage in Stokes IQUV would be thus a breathtaking 400-1800\,nm at $R=100,000$.


\begin{discussion}

\discuss{Cortes}{Science case for chirality in molecules at the 900 GHz range that ALMA could sense in polarimetry?}

\discuss{Strassmeier}{To my knowledge there are no laboratory measurements or simulations of circularly polarized transmissions or reflections from, say, leaves or any kind of cyanobacteria at such a wavelength. }

\discuss{Majallos}{As you pointed out, high accuracy will be needed and, hence, a control of systematics expected. What are you supposing as you planning?}

\discuss{Strassmeier}{The polarimeter itself is of a dual-beam design. Thus, we measure all Stokes parameters differentially, just like in classical differential photometry. If a systematic error applies to both beams, e.g. aging of the mirror coatings, they principally cancel out. The dual-beam design also allows swapping the beams and redoing the measurement. In practice the limitation is the change of the seeing while you repeat the first measurement. }

\end{discussion}


\begin{thebibliography}{}

\bibitem[2016]{angl}
Anglada-Escud\'e, G., Amado, P., Barnes, J. et al. 2016, Nature, 536, 437

\bibitem[2017]{bellu}
Belluzzi, L., Trujillo Bueno, J., \& Landi Degl Innocenti, E. 2017, ApJ, 666, 588

\bibitem[2016]{boy}
Boyajian, T., LaCourse, D., Rappaport, S., et al. 2016, MNRAS, 457, 3988

\bibitem[2012]{car:str}
Carroll, T. A., Strassmeier, K. G., Rice, J. B., \& K\"unstler, A. 2012, A\&A, 548, A95

\bibitem[2016]{divarano}
Di Varano, I., Strassmeier, K. G., Woche, M., \& Laux, U. 2016, SPIE 10012, 1001208

\bibitem[2018]{div}
Di Varano, I., Woche, M., Strassmeier, K. G., et al. 2018, SPIE Austin, in press

\bibitem[2006]{don:for}
Donati, J.-F.,  Forveille, T., Collier Cameron, A. et al. 2006, Science, 311, 633

\bibitem[1997]{don:bro}
Donati, J.-F. \& Brown, S. F. 1997, A\&A, 326, 1135

\bibitem[2017]{emde}
Emde, C., Buras-Schnell, R., Sterzik, M., \& Bagnulo, S. 2017, A\&A, 605, A2

\bibitem[2005]{gan}
Gandorfer, A. 2005, The Second Solar Spectrum, Vol. III, Zurich Hochschulverlag

\bibitem[2018]{garcia}
Garcia Munoz, A. 2018, ApJ, 854, 108

\bibitem[2016]{gill}
Gillon, M., Jehin, E., Lederer, S. et al. 2016, Nature, 533, 221

\bibitem[2017]{kuz}
Kuzmychov, O., Berdyugina, S., \& Harrington, D. 2017, ApJ, 847, 60

\bibitem[2018]{marconi}
Marconi, A., et al. 2018, SPIE Austin, in press

\bibitem[2015]{mar15}
Martins, J., Santos, N., Figueira, P. et al. 2015, A\&A, 576, A134

\bibitem[2018]{mar18}
Martins, J., Figueira, P., Santos, N., et al. 2018, MNRAS, 478, 5240

\bibitem[2014]{ovelar}
Ovelar, M. de Juan, Snik, F., Keller, C. U., \& Venema, L. 2014, A\&A, 562, A8

\bibitem[2002]{pis:koc}
Piskunov, N. E. \& Kochukhov, O. 2002, A\&A, 381, 736

\bibitem[2009]{rein}
Reiners, A., Basri, G., \& Christensen, U. 2009, ApJ, 697, 373

\bibitem[2014]{rob:enn}
Robinson, T. D., Ennico, K., Meadows, V. et al. 2014, ApJ, 787, 716

\bibitem[2018]{seng}
Sengupta, S. 2018, ApJ, 861, 41

\bibitem[2016]{sing}
Sing, D., Fortney, J., Nikolov, N. et al. 2016, Nature, 529, 59

\bibitem[2013]{snell}
Snellen, I., de Kok, R. J., le Poole, R., Brogi, M., \& Birkby, J. 2013, ApJ, 764, 182

\bibitem[2009]{snell09}
Snellen, I.,  de Mooij, E., \& Albrecht, S. 2009, Nature, 459, 543

\bibitem[2009]{spar}
Sparks, W. B.,  Hough, J., Germer, T. et al. 2009, PNAS, 106, 7816


\bibitem[2008]{stam08}
Stam, D. M. 2008, A\&A, 482, 989

\bibitem[2011]{ste11}
Stenflo, J. 2011, ASPC 437, 3

\bibitem[1997]{ste:kel}
Stenflo, J., \& Keller, C. U. 1997, A\&A, 321, 927

\bibitem[2012]{sterz}
Sterzik, M., Bagnulo, S., \& Palle, E. 2012, Nature, 483, 64

\bibitem[2017]{stol}
Stolker, T., Min, M., Stam, D. M., et al. 2017, A\&A, 607, A42

\bibitem[2012]{str12}
Strassmeier, K. G., DiVarano, I., Ilyin, I., et al. 2012, SPIE 8444, 35

\bibitem[2015]{pepsi}
Strassmeier, K. G., Ilyin, I., J\"arvinen, A. et al. 2015, AN, 336, 324

\bibitem[2018]{woche}
Woche, M., Di Varano, I., Strassmeier, K. G. et al. 2018, SPIE, Austin, in press

\bibitem[2010]{zugg}
Zugger, M., Kasting, J., Williams, D., Kane, T., \& Philbrick, C. 2010, ApJ, 723, 1168

\end{thebibliography}
\end{document}